\documentclass[12pt,a4paper]{article}
\pdfoutput=1
\usepackage{cite,color,graphics,amsmath,epsfig,rotating}
\usepackage{latexsym}
\usepackage{amssymb}

\usepackage{graphicx}
\usepackage{subfig}
\usepackage{psfrag}
\usepackage{mathrsfs}
\usepackage{url}
\usepackage[toc,page]{appendix}
\ifx\pdfoutput\undefined
\usepackage[bookmarks]{hyperref}	
\else
\usepackage{hyperref}	
\fi
\hypersetup{colorlinks=true,linktocpage,bookmarksopen,bookmarksnumbered,citecolor=azur,
linkcolor=green1,pdfstartview=FitH,urlcolor=darkred,breaklinks=true,
pdftitle={Collider limits on new physics within micrOMEGAs4.3},
pdfauthor={D. Barducci, G. B\'elanger, J. Bernon, F. Boudjema, J. Da Silva, S. Kraml, U. Laa, A. Pukhov},
pdfsubject={paper},
pdfkeywords ={beyond the standard model, new physics, dark matter, Higgs, LHC}}

\usepackage{multicol}
\usepackage{color}
\definecolor{azur}{rgb}{0.118,0.498,0.796}
\definecolor{darkred}{cmyk}{0,1,1,0.4}
\definecolor{green1}{rgb}{0.21,0.6,0.32}
\def\mhref#1{\href{mailto:#1}{#1}}		

\usepackage{marginnote}

\usepackage{eurosym}
\begin{document}

\setlength{\unitlength}{1mm}
\renewcommand{\arraystretch}{1.4}
\newcommand{\comment}[1]{}


\def\micro{{\tt micrOMEGAs}}
\def\microvn{{\tt micrOMEGAs\,4.3}}
\def\wimpsim{{\tt WimpSim}}
\def\pppc{{\tt }PPPC4DM$\nu$}
\def\pppcold{{\tt }DM$\nu$}
\def\chep{{\tt CalcHEP}}
\def\lhep{{\tt LanHEP}}
\def\darksusy{{\tt DarkSUSY}}
\def\smodels{{\tt SModelS}}
\def\madanalysis{{\tt MadAnalysis\,5}}
\def\checkmate{{\tt CheckMate}}
\def\nmssmtools{{\tt NMSSMTools}}
\def\fastlim{{\tt Fastlim}}
\def\xqcat{{\tt XQCAT}}
\def\lilith{{\tt Lilith}}
\def\HB{{\tt HiggsBounds}}
\def\HS{{\tt HiggsSignals}}
\def\suspect{{\tt SuSpect}}
\def\neuto{\tilde\chi^0_1}
\def\neuti{\tilde\chi^0_i}
\def\neutt{\tilde\chi^0_2}
\def\neuth{\tilde\chi^0_3}

\def\ie{{\it i.e.}}
\def\eg{{\it e.g.}}

\def\br{{\rm BR}}

\def\ra{\rightarrow}
\def\Omg{\Omega h^2}
\def\sip{\sigma^{SI}_{\chi p}}
\newcommand{\scs}{\scriptscriptstyle}
\def\simleq{\stackrel{<}{\scs \sim}}
\def\simgeq{\stackrel{>}{\scs \sim}}
\newcommand{\com}{\textcolor{green}}
\newcommand{\combis}{\textcolor{magenta}}

\newcommand{\ablabels}[3]{
  \begin{picture}(100,0)\setlength{\unitlength}{1mm}
    \put(#1,#3){\bf (a)}
    \put(#2,#3){\bf (b)}
  \end{picture}\\[-8mm]
}

\begin{titlepage}
\begin{center}

\vspace*{1.6cm}
{\Large\bf  Collider limits on new physics within micrOMEGAs$\_$4.3} 

\vspace*{1cm}\renewcommand{\thefootnote}{\fnsymbol{footnote}}

{\large 
D.~Barducci$^{1}$\footnote[1]{Email: \mhref{barducci@lapth.cnrs.fr}},
G.~B\'elanger$^{1}$\footnote[2]{Email: \mhref{belanger@lapth.cnrs.fr}},
J.~Bernon$^{2}$\footnote[3]{Email: \mhref{bernon@lpsc.in2p3.fr}},
F.~Boudjema$^{1}$\footnote[4]{Email: \mhref{boudjema@lapth.cnrs.fr}},
J.~Da Silva$^{1,2}$\footnote[5]{Email: \mhref{jonathan.dasilva@lapth.cnrs.fr}}, 
S.~Kraml$^{2}$\footnote[6]{Email: \mhref{sabine.kraml@lpsc.in2p3.fr}},
U.~Laa$^{1,2}$\footnote[7]{Email: \mhref{ursula.laa@lpsc.in2p3.fr}},
A.~Pukhov$^{3}$\footnote[8]{Email: \mhref{pukhov@lapth.cnrs.fr}},
} 

\renewcommand{\thefootnote}{\arabic{footnote}}

\vspace*{1cm} 
{\normalsize \it 
$^1\,$\href{http://lapth.cnrs.fr}{LAPTH}, Universit\'e Savoie Mont Blanc, CNRS, B.P.110,\\ F-74941 Annecy-le-Vieux Cedex, France\\[2mm]
$^2\,$\href{http://lpsc.in2p3.fr}{Laboratoire de Physique Subatomique et de Cosmologie}, Universit\'e Grenoble-Alpes,\\ CNRS/IN2P3, 53 Avenue des Martyrs, F-38026 Grenoble,  France\\[2mm]
$^3\,$\href{http://theory.sinp.msu.ru}{Skobeltsyn Institute of Nuclear Physics}, Moscow State University,\\ Moscow 119992, Russia\\[2mm]
}

\vspace{1cm}

\begin{abstract}
Results from the LHC put severe constraints on models of new physics. This includes constraints on the Higgs sector from the precise measurement of the mass and couplings of the 125~GeV Higgs boson, as well as limits from searches for other new particles. We present the procedure to use these constraints in \micro\  by interfacing it  to the external codes \lilith, \HS, \HB\ and \smodels.  A few dedicated modules are also provided. With these new features,  \verb|micrOMEGAs_4.3.4|  provides a generic framework for  
evaluating dark matter observables together with collider and non-collider constraints. 
\end{abstract}

\end{center}
\end{titlepage}

\tableofcontents

\section{Introduction}
A plethora of particle dark matter candidates have been proposed in the last decades to explain the astrophysical and cosmological observations via extensions of the Standard Model~(SM). 
In particular, new weakly interacting particles have attracted a lot of attention as natural cold dark matter candidates, see \eg~\cite{Bertone:2004pz,Bertone:2010zza} for comprehensive reviews. 
The list ranges from the well studied neutralino in the Minimal Supersymmetric Standard Model (MSSM)~\cite{Jungman:1995df,Griest:2000kj} to new fermionic or scalar states in a dark sector which communicates with the SM through some portal interactions~\cite{McDonald:1993ex,Patt:2006fw}. 
The code \micro~\cite{Belanger:2001fz,Belanger:2006is,Belanger:2008sj,Belanger:2010gh,Belanger:2013oya,Belanger:2014vza} was developed to compute the properties and signatures of a dark matter candidate in a generic model of  
new physics. Furthermore, through its link to CalcHEP~\cite{Pukhov:2004ca,Belyaev:2012qa}, observables at colliders can be computed. The code is widely used to compute dark matter observables and constraints in extensions of the SM.

With the extremely successful operation of the LHC and its experiments, which so far has led to the discovery of the Higgs boson~\cite{Aad:2012tfa,Chatrchyan:2012xdj} and to a large number of constraints on the signals of other new particles (including dark matter) in a vast variety of channels~\cite{atlas-susy,atlas-exotic,cms-susy,cms-exotic,cms-b2g}, 
checking the compatibility of a dark matter model with experimental observations  is becoming an increasingly complex and  time consuming task. This is  especially true in the framework of new physics models with a number of new particles and signatures at the TeV scale. 
Dedicated public codes have been developed to meet this challenge:  
\lilith~\cite{Bernon:2015hsa},  \HS~\cite{Bechtle:2013xfa} and \HB~\cite{Bechtle:2011sb,Bechtle:2013wla} to check Higgs sector constraints, as well as  
\checkmate~\cite{Drees:2013wra,Kim:2015wza}, \madanalysis~\cite{Dumont:2014tja,Conte:2014zja},  
\smodels~\cite{Kraml:2013mwa,Kraml:2014sna,Ambrogi:2017neo}, \fastlim~\cite{Papucci:2014rja}, \xqcat~\cite{Barducci:2014gna} 
and {\tt SUSY-AI}~\cite{Caron:2016hib}  for testing the limits from searches for supersymmetry (SUSY) or some other new physics. 
It is the philosophy of \micro\ to rely on other public codes when relevant. This has the advantage of providing a well-tested framework without duplicating work. 
We have therefore designed interfaces to some of the above-mentioned codes to confront the predictions of dark matter 
models to  the LHC results. 

In this paper we describe the facilities provided in \microvn\ to check the collider constraints on the dark matter models provided with the public distribution. These range from in-house routines designed for imposing LEP constraints or limits on a new $Z'$ at the LHC to interfaces to more complete codes that incorporate limits derived at the LHC Run~1. 
These include \lilith~\cite{Bernon:2015hsa} and \HS~\cite{Bechtle:2013xfa} that check the compatibility to measurements of the Higgs boson at 125 GeV, \HB~\cite{Bechtle:2011sb,Bechtle:2013wla} that provides model-independent limits on additional neutral and charged Higgs states, and \smodels~\cite{Kraml:2013mwa,Kraml:2014sna,Ambrogi:2017neo} that 
checks the LHC limits on the production of new odd particles in the context of Simplified Model Spectra (SMS) constraints. 
For each of these facilities, we explain the functions that 
are provided in \micro\ to impose constraints, their input format and the interpretation of the output.

The codes chosen for the interface with \micro\ are not only fast  and therefore suitable for large scans of parameter space, they are 
also quite generic and thus applicable to a wide range of models. Of course some limitations apply. For example, the Higgs 
signal strength approach in \lilith\ and \HS\ is not adapted for models where new structures in the Higgs vertices or new 
Higgs production modes appear. Moreover, the SMS results used in \smodels\  might not always be directly applicable to cases with a different spin structure than considered by the experiments. The user is advised to keep these limitations in mind for the proper usage of these codes in \micro.

The paper is organised as follows. 
Section~\ref{sec:Higgs} deals with the tools for testing the Higgs sector: \lilith, \HS\ and \HB. 
Section~\ref{sec:susy} describes the interface to \smodels\ as well as the routines to test LEP limits and the routine to constrain a $Z'$. An outlook for a new routine evaluating constraints from monojet searches is also given.
Installation instructions as well as a sample output are given in Section~\ref{sec:install}.
Section~\ref{sec:Conclusion} contains our conclusions.

\section{Higgs-sector constraints} 
\label{sec:Higgs}
Constraints originating from the measurements of signal strengths of the 125 GeV state at LHC can be obtained by means of \lilith~\cite{Bernon:2015hsa} or \HS~\cite{Bechtle:2013xfa} while constraints on additional Higgs states are accessible through \HB~\cite{Bechtle:2013wla}.

\subsection{Lilith}\label{sec:lilith}

\lilith~\cite{Bernon:2015hsa} is  a light and very fast {\tt Python} library that can be used to derive constraints on the parameter space of new physics scenarios. A global likelihood function $\mathcal{L}$ is constructed from the latest ATLAS and CMS results. The \lilith\ inputs are the set of reduced couplings of the 125 GeV state, \ie, couplings normalized by the SM ones, and possible invisible or undetected branching ratios.  Note that \lilith\ can test Higgs bosons with masses within the $[123, 128]$~GeV interval, a warning will be issued if no such state can be found. In the case where two or more states have masses within this interval, their signal strengths will be summed incoherently and an effective Higgs state will be tested against the LHC measurements.

For a given parameter space point $\mathcal{P}$ of a new physics scenario, from which reduced couplings and new branching ratios can be obtained, \lilith\ returns ($-2$ times) the log-likelihood evaluated at this point, $-2\log\mathcal{L}(\mathcal{P})$, as,
\begin{equation}
-2\log\mathcal{L}(\mathcal{P}) \equiv \sum_{i=1}^m -2\log\mathcal{L}^{exp}_i(\mathcal{P}),
\end{equation}
where $i$ indexes the $m$ measurements used to construct the global likelihood function $\mathcal{L}$ and $\mathcal{L}^{exp}_i$ are the individual experimental likelihood functions.
If all the $\mathcal{L}^{exp}_i$ functions are Gaussian, $-2\log\mathcal{L}(\mathcal{P})$ is identified as a chi-squared ($\chi^2$). In practice, most of the measurements are well within the Gaussian regime, with the notable exception of the $H\to ZZ^*$ measurements suffering from a small number of observed events. One could assess the statistical compatibility of $\mathcal{P}$ with the combined set of experimental results by computing a p-value based on a $\chi^2$ distribution with $n^{exp}\equiv\sum_{i=1}^m n_i$ degrees of freedom, referred to as the \textit{number of experimental degrees of freedom}, where $n_i$ is the number of variables of the likelihood function $\mathcal{L}_i^{exp}$. The number of experimental degrees of freedom is completely determined by the database of experimental results used within {\tt Lilith} to construct the global Likelihood function $\mathcal{L}$. So far, experimental collaborations have delivered likelihood functions depending on $n_i=1,2$ variables (mostly identified as signal strengths), that can often be approximated by 1,\,2-dimensional Gaussian functions, respectively.\footnote{Note that for 2-dimensional likelihood functions, the correlation between the two variables as determined by the experimental collaborations are completely accounted for.} This statistical interpretation is used by default within \micro. We detail in the following the calculation of the p-value.

In general, the p-value is computed from a $\chi^2$ distribution with $ndf$ degrees of freedom, for an observed value of
\begin{equation}
   \label{deltaL}
   \Delta (-2\log\mathcal{L})\equiv -2\log\mathcal{L}(\mathcal{P}) - \left( -2\log\mathcal{L}(\text{ref}) \right)\,,
\end{equation}
where $\mathcal{L}(\text{ref})$ is a reference likelihood point.
Explicitly, the p-value $p$ is obtained as
\begin{equation}
   \label{pval}
   p=\int_{\Delta (-2\log\mathcal{L})}^{+\infty} \chi^2(x;ndf) dx\ ,
\end{equation}
where $\chi^2(x;ndf)$ is the probability density function of a $\chi^2$ distribution with $ndf$ degrees of freedom and
\begin{equation}
\label{ndf}
ndf \equiv n^{exp}-n^{par}\,,
\end{equation}
where $n^{exp}$ are the experimental degrees of freedom and $n^{par}$ the number of relevant free parameters.\footnote{\label{footnote_ndf}One should keep in mind, however, that in the case of a non-linear model, as obtained when parametrizing signal strengths by reduced couplings for instance, Eq.~\eqref{ndf} may in fact be a very crude estimate of the real number of degrees of freedom; see Ref.~\cite{2010arXiv1012.3754A} for a detailed discussion of this problem.}  In \micro, the two parameters $-2\log\mathcal{L}(\text{ref})$ (a non-negative real) and $n^{par}$ (a~non-negative integer) are free parameters and are dubbed \verb|m2logL_reference| and \verb|n_par|, respectively.

By default in \micro,  both \verb|n_par| and \verb|m2logL_reference| are set to 0, corresponding to a p-value $p^{def}$ computed with $n^{exp}$ degrees of freedom and $-2\log\mathcal{L}(\text{ref})=0$.
The default p-value is thus obtained~as
\begin{equation}
   \label{pval2}
   p^{def}=\int_{-2\log\mathcal{L}(\mathcal{P})}^{+\infty} \chi^2(x;n^{exp}) dx\ .
\end{equation}

The file \verb|include/Lilith.inc| (or \verb|Lilith.inc_f|) contains the instructions to launch \lilith\ using a system call. 
The input file  \verb|Lilith_in.xml| for \lilith\ is created automatically via the command\\

\verb|LilithMDL("Lilith_in.xml")|\\

\noindent 
which returns the number of neutral Higgs particles and the file containing the list of reduced Higgs couplings as well as the branching ratios of Higgs decays to invisible, $\br_{\rm inv}$, and to other non-SM final states, $\br_{\rm undetected}=1-\br_{\rm inv}-\sum \br(H\rightarrow {\rm SM \,SM})$.\footnote{\label{BRinvCaveat}By default, the automatic generation of the input file assumes that only DM contributes to the invisible width. }
Note that the reduced couplings of the 125~GeV Higgs are defined for all the models provided with \micro\ with the exception of the $Z\gamma$ coupling. The latter is computed within \lilith\ assuming that only SM particles run in the loop. 
The command \verb|LilithMDL| is model dependent but can be easily adapted for user-implemented models. In this case, the reduced couplings have to be defined in \verb|lib/lilith.c| of the new model, including the effective Lagrangian for the loop-induced couplings. Alternatively  \micro\ contains a new option to generate automatically  the input file.  To use this option, it suffices to replace the above call to \verb|LilithMDL| by\\

\verb|LilithMO("Lilith_in.xml")|\\

\noindent
The functionality of \verb|LilithMO| is described in Section~\ref{sec:auto}.

The input parameters \verb|n_par| and \verb|m2logL_reference| are defined in the \verb|main.c| file of each model.
The {\tt SLHA} output file\footnote{We use here the name  SLHA file to designate any file that has the  {\tt BLOCK} structure  defined in SLHA~\cite{Skands:2003cj,Allanach:2008qq} even if the model is  non supersymmetric.}, \verb|Lilith_out.slha|, consists in six entries which are respectively $-2\log\mathcal{L}(\mathcal{P})$, $n^{exp}$, $-2\log\mathcal{L}(\mathcal{\text{ref}})$, $ndf$, $p$ and the database version.
For instance, for the \verb|mssmh.par| point of the MSSM model, one obtains $-2\log\mathcal{L}(\mathcal{P})=28.1285$ and $n^{exp}=38$, leading to $p^{def}=0.879$.  
In this case the database \verb|DB_15.09| was used and the following {\tt SLHA} output was generated, 
\begin{verbatim} 
BLOCK         LilithResults
  0           28.1285          # -2*LogL
  1           38               # exp_ndf
  2           0.0              # -2*LogL_ref
  3           38               # fit_ndf
  4           0.879            # pvalue
  5           15.09            # database version
\end{verbatim}
The user is free to use the computed p-value at her/his will. For instance, one could flag or exclude points with too low p-value. In this context, a point with a p-value smaller than $0.3173$, $0.0455$, $0.0027$ could be excluded at more than the 1$\sigma$, 2$\sigma$, 3$\sigma$ levels, respectively.

For generic values of \verb|m2logL_reference|, Eq.~\eqref{pval} actually describes the p-value for a likelihood-ratio test in the asymptotic limit. Indeed, according to Wilk's theorem, such a statistical test is asymptotically distributed as a $\chi^2$ distribution, with $ndf\equiv n^{par}\neq 0$ standing for the number of free parameters controlling the Higgs sector of the scenario under study. Such a statistical procedure is well suited to perform a fit of the model, in which case one would take the best-fit point of the model, \ie, the point for which the global minimum of $-2\log\mathcal{L}$ is attained, as the reference in Eq.~\eqref{deltaL}. 
The best-fit point can be found by a preliminary scan or using a numerical minimization method for instance. Here, if \verb|m2logL_reference| is set to $-1$, the SM likelihood will be used as the reference point. As long as the SM is a good description of the 125~GeV state properties,%
\footnote{The SM provides an excellent fit to the current Higgs measurements; with \lilith\ {\tt DB\_15.09}, which contains all the Run~1 results,  we find $-2\log\mathcal{L}({\rm SM})=25.9482$ with $n^{exp}=38$, that is a $p$-value of $p^{def}=0.9312$.} 
the SM likelihood should provide a good approximation of the best-fit point of any model with a SM limit. In the case of a large number of parameters, it may be justified to regard all reduced couplings as independent from each other, and thus identify $n^{par}$ as the number of reduced couplings that receive new physics contributions.

In Table~\ref{lilith_parameters}, we summarize the roles of \verb|m2logL_reference| and \verb|n_par|, the two free parameters controlling the p-value calculation detailed in Eq.~\eqref{pval}.
By default, both \verb|n_par| and \verb|m2logL_reference| are set to 0, which corresponds to the first line of Table~\ref{lilith_parameters} with $n^{par}=0$.
If only \verb|n_par| is given a non-vanishing value, the p-value will be computed with a smaller number of degrees of freedom, as shown in the first line of the table.
If \verb|m2logL_reference| is set to a positive value, \verb|n_par| should also be set to a value different from 0 (it is interpreted as the number of parameters controlling the Higgs sector of the given scenario) and the p-value will be computed with modified $ndf$ and $\Delta (-2\log\mathcal{L})$ according to the second line of the table.
On the other hand, if \verb|m2logL_reference| is set to $-1$, the SM will be used as the reference point in $\Delta (-2\log\mathcal{L})$ (see Eq.~\eqref{deltaL}) as shown in the last line of Table~\ref{lilith_parameters}.
\begin{table}[t!]
\begin{center}
\begin{tabular}{|cc|cc|}
  \hline
  \begin{tabular}[x]{@{}c@{}} $-2\log\mathcal{L}(\text{ref})$\\(\verb|m2logL_reference|)\end{tabular}  &\begin{tabular}[x]{@{}c@{}}$n^{par}$\\ (\verb|n_par|)\end{tabular}  & $\Delta (-2\log\mathcal{L})$& $ndf$  \\
  \hline
   $-2\log\mathcal{L}(\text{ref})=0$ & $n^{par}\geq0$  & $-2\log\mathcal{L}(\mathcal{P})$ & $n^{exp}-n^{par}$ \\
  \hline
   $-2\log\mathcal{L}(\text{ref})>0$ &$n^{par}\geq1$ &  $-2\log\mathcal{L}(\mathcal{P})-\left(-2\log\mathcal{L}(\text{ref})\right)$ & $n^{par}$  \\
  \hline
   $-2\log\mathcal{L}(\text{ref})=-1$ &$n^{par}\geq1$  & $-2\log\mathcal{L}(\mathcal{P})-\left( -2\log\mathcal{L}(\text{SM}) \right)$ &  $n^{par}$ \\
  \hline
\end{tabular}
\caption{Roles of {\tt m2logL\_reference} and  {\tt n\_par} in the {\tt Lilith} p-value calculation within {\tt micrOMEGAs}. The two left-most columns indicate the various possible inputs, while the two right-most columns give the corresponding parameters entering the p-value calculation of Eq.~\eqref{pval}.}
\label{lilith_parameters}
\end{center}
\end{table}

\subsection{HiggsBounds and HiggsSignals}

Constraints on the properties of the 125 GeV Higgs boson can also  be obtained with \HS.
Moreover exclusion limits provided by the experimental LHC and Tevatron collaborations on additional Higgs bosons are obtained through an interface to \HB~\cite{Bechtle:2011sb}.
The interface to these codes has been updated with respect to previous versions, moreover theses codes are no longer distributed with \micro\ but are downloaded when required. 
The file {\tt include/hBandS.inc}  contains the instructions to call both \HB\ and \HS, see~\cite{Bechtle:2013wla} for more details on the input options.
 
In~\micro, the same SLHA file is used  as an input to both of these codes,   it relies on  the effective coupling option for specifying the input.  For models distributed with~\micro, the  function\\

\verb| hbBlocksMDL("HB.in",&NchHiggs) | \\

\noindent
can be used to  write at the end of the SLHA file, \verb|HB.in|, new blocks that contain the Higgs masses, the effective couplings of the Higgses normalized to the SM ones, 
including the reduced couplings squared to  $\gamma\gamma,\gamma Z, gg$
as well as all Higgs total widths  and branching ratios and the  top decay width.
A  reference value is defined for the couplings that do not exist in the Standard Model. This routine returns the  number of neutral Higgses, and \verb|NchHiggs| gives the number of charged Higgs particles. 
For models not included in the \micro\ distribution, the user can  rewrite the function {\tt hbBlocksMDL} located in the \verb|lib| directory of the model or provide
an SLHA file containing the reduced couplings squared. Moreover, the number of neutral and charged Higgs states must be provided, 
 and the theoretical uncertainty on the masses of these particles must  be specified in the \verb|BLOCK DMASS|. Alternatively,  the user can use the new option to generate automatically  the input file  by  calling\\

\verb| hbBlocksMO("HB.in",&NchHiggs) | \\

\noindent
The content of this function is described in Section~\ref{sec:auto}. 
The complete outputs of \HB\ and \HS\ are stored in the files  \verb|HB.out| and \verb|HS.out| respectively and can be accessed  and read  by the user using the \verb|slhaval| function~\cite{Belanger:2014hqa}.
The screen output of \micro\ contains the following information\\
 
\verb| HB(version number): result  obsratio  channel |\\
 
\noindent
where \verb|result| = $0,1,-1$ denotes respectively whether a parameter point is excluded at 95\% CL, not excluded, or invalid; \verb|obsratio| gives the ratio of the theoretical expectation relative to the observed value  for the most constraining channel specified in \verb|channel|. The \HS\ output displayed on the screen is simply \\

\verb| HS(version number): Nobservables chi^2  pval|\\

\noindent
where \verb|Nobservables| gives the number of observables used in the fit, \verb|chi^2| the associated $\chi^2$ and \verb|pval| the p-value. 

The interpretation of these values is left to the user. Note here that \verb|pval|\  is determined in the same way as the default p-value in \lilith, see explanations to Eq.~\eqref{pval2}, unless the user explicitly provides a non-zero value for the number of parameters in the \HS\ input file \verb|HS.in|, in which case $ndf = nobs - npar$ is used. 
This gives a hypothesis test of how likely the model point explains the data, under the assumptions that all measurements (in all $\sim100$ sub-categories considered in \HS) are nicely Gaussian; see however footnote~\ref{footnote_ndf} for caution about this definition of $ndf$.

To determine how much a model point is favoured or disfavoured over the SM, one can use $\Delta\chi^2$ to compute the likelihood ratio in the asymptotic limit as explained in the next-to-last paragraph of Section~\ref{sec:lilith}. We remind the reader that in this case the relevant number of degrees of freedom is the number of parameters (\ie\ number of Higgs couplings receiving new physics contributions) instead of the number of observables. Identifying the SM as the best fit, the likelihood ratio simplifies to   
$p(\Delta\chi^2;\,npar)$ with $\Delta\chi^2=\chi^2-\chi^2({\rm SM})$.

\subsection{Automatic generation of interface files}
\label{sec:auto}

The functions {\tt LiLithMO} and {\tt hbBlocksMO} provided for generating automatically the input files for {\tt Lilith}  and {\tt HiggsBounds}/{\tt HiggsSignals} 
for new models contain two routines  that allow  to extract the couplings contained in the model file, {\tt lgrng1.mdl}.  
These routines are described below.  We stress however that they do not have to be called explicitly by the user. 
The first routine returns a description of a given vertex. The format used is \\

\verb|lVert* vv=getLagrVertex(name1,name2,name3,name4);|\\

\noindent
where {\it name1,..,name4 } are the names of the particles included  in the vertex; for vertices with three particles, {\it name4} should be replaced by
{\tt NULL}.  The return parameter {\tt vv} is the memory address of a structure which contains information about the vertex:
\begin{itemize}
\item \verb|vv->GGpower    | - power of strong coupling included in vertex
\item \verb|vv->nTerms     | - number of different  Lorentz structures in vertex
\item \verb|vv->SymbVert[i]| - text form of Lorentz structures $i\in [0,nTerms]$  
\end{itemize}
The second routine allows to obtain the numerical coefficients corresponding to  each  Lorentz structure. The command is\\

\verb|getNumCoeff(vv,coeff)| \\

\noindent
with {\tt coeff[i]} the numerical coefficient for {\tt SymbVert[i]}.
Note that the strong coupling is factored out of the coefficients. 
For example, for the standard three gauge bosons interaction the  {\tt SymbVert}  array and coefficients are 
\begin{eqnarray}
\nonumber
   {\rm SymbVert} &= \{& \verb|p1.m1|*\verb|m3.m2|,\;  \verb|p2.m1|*\verb|m3.m2|,\;  \verb|p1.m2|*\verb|m3.m1|,\\
            &&  \verb|p2.m2|*\verb|m3.m1|,\; \verb|m2.m1|*\verb|p1.m3|,\; \verb|m2.m1|*\verb|p2.m3|\;\} \\
   {\rm coeff}    &=\{& x, 2x, -2x ,-x,x,x\; \}
\end{eqnarray}  
where $x$ is the  electromagnetic coupling $e$ for  $W^+W^-\gamma$ and
$ x=\frac{e}{\tan{\Theta_W}}$ for $W^+W^-Z$. Using  these two vertices, \micro\ defines  the electroweak parameters required for the computation 
of the  reduced  Higgs coupling in  the  model.

All the QCD-neutral scalars belonging to the even sector (not designated with \verb|~|)  are considered as  Higgs particles. For each of these, \micro\ calculates  the couplings to  SM fermions and massive  bosons and  writes down into the interface file the ratio of these couplings to the  corresponding  SM Higgs coupling. Note that the  couplings of the Higgs to   SM  fermions can significantly depend on the QCD scale; \micro\ assumes that the quark masses entering the vertices are obtained at the same scale in both  the
SM and  the new model, thus the scale dependence in the  reduced couplings to fermions disappears. 
The loop-induced  couplings of the Higgs to  gluons and photons are calculated by the {\tt LiLithMO/hbBlocksMO}  routines whether or not the $Hgg$ and $H\gamma\gamma$ vertices are already implemented in the Lagrangian. This includes NLO-QCD corrections and is performed as described in~\cite{Belanger:2013oya,Djouadi:1997yw}.

In addition to the reduced couplings, {\tt Lilith} requires the branching ratios of the Higgs decays to invisibles and to other non-SM particles, while  {\tt HiggsBounds/HiggsSignals}  requires all Higgs branching ratios as well as the total widths. These are also written automatically in the interface file (note however the caveat in footnote~\ref{BRinvCaveat}). When the Higgs widths and branching ratios are provided in an SLHA file, these values will be used. Otherwise {\tt LiLithMO/hbBlocksMO}  check  the existence
of $H\rightarrow gg$  and $H\rightarrow \gamma\gamma$  in the table of decays generated from the model Lagrangian. If found, the branching ratios and total widths are written in the  interface file without comparing  with the internal calculations. If not found, then {\tt LiLithMO/hbBlocksMO} add these channels and  recompute the total widths and all branching ratios.

\section{Collider limits on new particles} 
\label{sec:susy}

\subsection{LEP limits}\label{sec:lep}

Generic limits from LEP  on charged supersymmetric particles have been implemented 
since the first version of \micro~\cite{Belanger:2001fz}. 
The relevant function is 
\begin{quote}
  \verb|masslimits()|
\end{quote}
which returns a value greater than 1 and prints a Warning when the mass of at least one of the new particles conflicts with a direct limit on sparticle masses from  LEP.
The constraints on the  Higgs sector from LEP are not implemented in this function. 

The evaluation of two additional constraints are now provided. 
The first one is the limit on the invisible width of the $Z$ boson, $\Gamma_{inv}(Z)<0.5$~MeV~\cite{Freitas:2014hra}, 
which is relevant when the DM candidate is lighter than $M_Z/2$. This can be checked by calling the function 
\begin{quote}
  \verb|Zinvisible()|
\end{quote}
which returns 1 and prints a Warning when $\Gamma_{inv}(Z)>0.5$~MeV. This function can be used in any model with one or two DM candidates where the $Z$ boson is defined by its PDG code (23). 
The second is the upper limit \cite{Abbiendi:2003sc} on the cross section for the production of neutralinos 
$\sigma(e^+e^-\rightarrow \neuto\neuti)$, $i\neq 1$, when the heavier neutralino decays into quark pairs 
and the LSP, $\neuti \rightarrow \neuto q\bar{q}$. The relevant function is  
\begin{quote}
  \verb|LspNlsp_LEP()|
\end{quote}
which returns $\sigma \times BR = \sum_i \sigma(e^+e^-\rightarrow \neuto\neuti)\times {\rm BR}(\neuti \rightarrow \neuto q\bar{q})$ in pb as well as a flag greater than one if $\sigma \times BR>0.1(0.5)$~pb if $m_{\rm NLSP}> (<)100$~GeV \cite{Abbiendi:2003sc}. 
This function can also be applied for non-SUSY models which feature the same signature, in this case the function will compute the cross section for production of the LSP and any other neutral particle from the odd sector which can decay into the LSP and a $Z$ boson.

\subsection{LHC limits from SModelS}

The new particles present in extensions of the SM with a $\mathbb{Z}_2$ symmetry  are strongly constrained by the LHC results.
In particular, LHC limits on new (odd) particles can be obtained using \smodels~\cite{Kraml:2013mwa,Kraml:2014sna,Ambrogi:2017neo}, 
a code which tests Beyond the Standard Model (BSM) predictions against Simplified Model Spectra (SMS) results from searches 
for R-parity conserving SUSY by ATLAS and CMS.
\smodels~v1.1.0 decomposes any BSM model featuring a $\mathbb{Z}_2$ symmetry into its SMS components 
using a generic procedure where each SMS is defined by the vertex structure and the SM final state particles; BSM particles are described only by their masses, production cross sections and branching ratios. 
The underlying assumption is that differences in the event kinematics (\eg\ from different production mechanisms or from the spin of the BSM particle) do not significantly affect the signal selection efficiencies. 
Within this assumption, \smodels\ can be used for any BSM model with a $\mathbb{Z}_2$ symmetry as long as all heavier odd particles decay promptly to the dark matter candidate.\footnote{Charged tracks may also be treated in an SMS context, see~\cite{Heisig:2015yla}, and will be available in future versions of \smodels.}
Note that due to the $\mathbb{Z}_2$ symmetry only pair production is considered, and missing transverse energy (MET) is always implied in the final state description.

\subsubsection{Input files}

\smodels\ needs three files:
\begin{itemize}
\item an SLHA-type input file,  
containing the mass
spectrum, decay tables\footnote{Note that all decay products in the decay table need to be on-shell.} and production cross sections for the parameter point under investigation;
\item {\tt particles.py} defining the particle content of the model, specifically which particles are even (``R-even'') 
and which ones are odd (``R-odd'') under the $\mathbb{Z}_2$ symmetry;
\item a file for setting the run parameters, {\tt parameters.ini}.
\end{itemize}
The first two are located in the same directory as {\tt main.c} and 
are automatically written by \micro\ by calling the function
\begin{quote}
  \verb|smodels(Pcm, nf, csMinFb, fileName, wrt)|
\end{quote}  
where \verb|Pcm| is the proton beam energy in GeV and \verb|nf| is the number of parton flavors used to compute the production cross sections of the $\mathbb{Z}_2$-odd particles. 
(Note that $u$, $d$, $\bar{u}$, $\bar{d}$  and gluons are always included  while  $s$, $c$, and $b$ quarks are included for ${\tt nf} = 3,4,5$ respectively.)
{\tt csMinFb} defines the minimum production cross section in pb for $\mathbb{Z}_2$-odd   particles; processes with lower cross sections are not added to the SLHA file passed to \smodels, here denoted by  {\tt fileName}. 
Finally, {\tt wrt} is a steering flag for the screen output; if ${\tt wrt} \ne 0$  the computed cross sections will be also written on the screen.

In the specific case of SUSY models, \smodels\ can be used to call {\tt nllFast}~\cite{nllfast} for the calculation of k-factors.  The higher-order (NLO+NLL) cross sections  for strong production processes are then added to the input file. 
For example the k-factors for 8 TeV cross sections can be added via the system call
\begin{verbatim}
  /runTools.py -particles ./ xseccomputer -p -s 8 -N -O -f fileName
\end{verbatim}
where {\tt fileName} is the name of the \smodels\ input file already used above. A file containting the instructions to call  \smodels\ can be found in \verb|micromegas_4.3.4/include/SMODELS.inc| (or \verb|SMODELS.inc_f|) where {\tt fileName = smodels.in}.

The SMS decomposition and confrontation against the LHC limits are also executed  
via a system call,
\begin{verbatim}
/runSModelS.py -f fileName -o ./ -p parameters.ini -particles ./ -v error
\end{verbatim}
\normalsize
where {\tt -o} sets the directory for the output file, {\tt -v} controls the level of smodels output (only error messages will be printed in this example)
and {\tt parameters.ini} is the file that can be used to set the run parameters (see below).
Running \smodels\ will produce two output files in the selected directory, a text summary in {\tt filename.smodels} and
an SLHA-type output in {\tt filename.smodelsslha}.
Note that a binary database is built when running \smodels\ for the first time.
This can take a few minutes, but needs to be done only once.

Before testing against LHC limits, \smodels\ applies consistency checks on the input point, verifying \eg\ that all decay tables are written in the input file and that all decays are kinematically allowed. 
It also checks whether there are long-lived charged particles and/or displaced vertices; in such a case the point is discarded and labelled as {\em not tested}. 
For all points passing the checks, \smodels\ proceeds to the decomposition into SMS components and matches them to the experimental upper limits in the database.

There are two types of results in the \smodels\ database.
Upper limit (UL) maps directly report an upper limit on the topology weight, computed as production cross
section times branching ratios, while efficiency maps (EM) report the signal selection efficiency.
For EM type results \smodels\ collects all contributions to one signal region and calculates the cross section rescaled by the appropriate efficiencies.
They can then be compared to an overall limit on the visible cross section.\footnote{The current database version {\tt 1.1.0patch1} contains results from 25 ATLAS and 23 CMS SUSY searches at 8 TeV, and results from 3 early 13 TeV searches. In addition, efficiency maps derived by the \fastlim\ collaboration~\cite{Papucci:2014rja} are also included in the database. The database can easily be updated with additional results independent of code updates~\cite{Ambrogi:2017neo}.}

For each matching result, \smodels\ reports an $R$ value, defined as the ratio of the predicted theory cross section and the corresponding experimental upper limit.
An  $R$ value larger than $1$ indicates that the point is excluded by the corresponding search.
Additional information (expected $R$ value, $\chi^2$ and likelihood computation) are available for EM type results.
In addition, \smodels\ returns information about important topologies for which no matching result exists.
These so-called ``missing topologies'' are specified in the bracket notation defined in~\cite{Kraml:2013mwa} and can be used to design new searches or Simplified Models that can constrain the scenario further.

Specific run parameters for \smodels\  can be set in a parameter file {\tt parameters.ini}.
A commented example can be found in {\tt Packages/smodels-v1.1.0patch1/parameters.ini}.
If no parameter file is specified, {\tt Packages/smodels-v1.1.0patch1/etc/parameters\_default.ini} is used.
Concretely, 
\begin{description}
\item[] {\tt doInvisible} and {\tt doCompress} are used for turning on/off invisible and mass compression (on by default), turning on the former entails the compression of vertices where all SM particles are invisible in the detector (\eg, neutrinos).
\item[] {\tt minmassgap} (default $5$ GeV) is the minimum mass gap for mass compression and 
\item[] {\tt sigmacut} (default $0.03$ fb) is the cutoff cross section for topologies to be considered in the decomposition. Note that this value is independent from  {\tt csMinFb} that is only used by \micro\ for writing the input file.
\item[] {\tt maxcond} (default 0.2) sets the maximum condition violation for a result to be considered.
Conditions are used when an experimental analysis combines final states with different selection efficiencies to evaluate a single upper limit (for example electron and muon final states). The condition is considered violated if the predicted composition differs from the one assumed in the limit setting procedure such that the constraint would be too strong. It is quantified as a relative difference from 0 (no violation) to~1 (condition maximally violated).
\end{description}
Additionally, the {\tt parameter.ini} file allows to select only specific results from the database, 
by specifying  {\tt analyses} or {\tt txnames}.
For detailed explanations of these functionalities see~
\cite{Ambrogi:2017neo}.

\subsubsection{Output format}

An SLHA-type output format was designed for the \smodels--\micro\ interface, and is written to {\tt filename.smodelsslha} (in the directory selected by the user).
This output consists of the blocks, {\tt SModelS\_Settings}, {\tt SModelS\_Exclusion} specifying the settings and constraints, and the blocks {\tt SModelS\_Missing\_Topos}, {\tt SModelS\_Outside\_Grid}, {\tt SModelS\_Long\_Cascade} and {\tt SModelS\_Asymmetric\_Branches} detailing information about the coverage by Simplified Models. 
Below we give a description of each block together with a sample output corresponding to the file {\tt mssm1.par} in the MSSM directory.

\begin{itemize}
\item {\tt SModelS\_Settings} lists the \smodels\ code and database versions as well as input parameters for the decomposition. For example:

\begin{verbatim}
BLOCK SModelS_Settings
 0 v1.1.0patch1            #SModelS version
 1 1.1.0patch1             #database version
 2 0.2                     #maximum condition violation
 3 1                       #compression (0 off, 1 on)
 4 5.0                     #minimum mass gap for mass compression [GeV]
 5 0.03                    #sigmacut [fb]
 \end{verbatim}

\item {\tt SModelS\_Exclusion} 
contains as the first line (the {\tt 0 0} entry) the status information if a point is excluded (1), not excluded (0),  or not tested ( $-1$). 
The latter can occur in scenarios with long-lived charged particles or in scenarios where no matching SMS results are found. 

If a point is excluded (status 1), this is followed by a list of all results with $R>1$, sorted by their $R$ values. 
For each of these results, the SMS topology identifier (entry 0) (so-called Tx-name, see~\cite{SMS_dictionary} for an explanation of the terminology), 
the $R$ value (entry 1), for efficiency maps  results the expected $R$ value (entry 2), a measure of condition violation (entry 3), and the analysis identifier (entry 4) are listed. Entries 5, 6 and 7 are relevant only for EM type results, and specify the
most sensitive signal region (used for limit setting), the $\chi^2$ and the likelihood value respectively.
If the point is not excluded (status 0), the result with the highest $R$ value is given instead to show whether a point is close to the exclusion limit or not. 

In the example below, obtained from {\tt mssm1.par}, the  highest $R$ values correspond to  a CMS supersymmetry search in the hadronic final states using $M_{T2}$~\cite{Khachatryan:2015vra} and a 
dijet+MET search constraining squark production, with $\tilde{q}\rightarrow q \tilde{\chi}^0_1$ obtained by ATLAS~\cite{Aad:2014wea}. 
Note that only the first part of the file is reproduced.

\begin{verbatim}
BLOCK SModelS_Exclusion
 0 0 1    #output status (-1 not tested, 0 not excluded, 1 excluded)
 1 0 T2                   #txname 
 1 1 6.161E+00            #r value
 1 2 N/A                  #expected r value
 1 3 0.00                 #condition violation
 1 4 CMS-SUS-13-019       #analysis
 1 5 (UL)                 #signal region 
 1 6 N/A                  #Chi2
 1 7 N/A                  #Likelihood

 2 0 T2                   #txname 
 2 1 6.156E+00            #r value
 2 2 7.077E+00            #expected r value
 2 3 0.00                 #condition violation
 2 4 ATLAS-SUSY-2013-02   #analysis
 2 5 SR4jl-               #signal region 
 2 6 2.298E+01            #Chi2
 2 7 3.402E-08            #Likelihood
\end{verbatim}

\item {\tt SModelS\_Missing\_Topos} lists up to ten missing topologies sorted by their weights ($ = \sigma\times {\rm BR}$).  Each entry consists of the line number, the $\sqrt{s}$ in TeV, the weight and a description of the topology in the \smodels\ bracket notation. Note that this information is useful mainly for points that are not excluded.

\begin{verbatim}
BLOCK SModelS_Missing_Topos #sqrts[TeV] weight[fb] description
 0 8  1.357E+03 [[[jet]],[[jet],[jet]]]
 1 8  3.382E+02 [[[b],[b]],[[jet]]]
 2 8  2.796E+02 [[],[[jet]]]
 3 8  2.532E+02 [[[b]],[[jet]]]
 4 8  2.510E+02 [[[jet]],[[t],[b],[W]]]
 5 8  2.274E+02 [[[b],[W]],[[b],[nu],[ta]]]
 6 8  1.090E+02 [[[jet]],[[jet],[jet],[jet]]]
 7 8  4.709E+01 [[[jet],[jet]],[[jet],[jet]]]
 8 8  4.231E+01 [[[jet]],[[jet],[W]]]
 9 8  3.197E+01 [[[jet]],[[jet],[W],[W]]]
\end{verbatim}
\end{itemize}

\noindent 
The blocks {\tt SModelS\_Outside\_Grid}, {\tt SModelS\_Long\_Cascade} and {\tt SModelS\_Asymmetric\_Branches} are similar to the {\tt SModelS\_Missing\_Topos} block; we refer the reader to~\cite{Ambrogi:2017neo} for details.

\subsubsection{Identification of the SM-like Higgs}

Several SUSY searches exploit decay channels of heavy neutralinos to a SM-like Higgs and the LSP. 
For example, the three leptons ATLAS search~\cite{Aad:2014nua} has an SMS interpretation for $\tilde{\chi}^{\pm}_1 \tilde{\chi}^{0}_2$ production, where $\tilde{\chi}^{\pm}_1 \rightarrow W \tilde{\chi}^{0}_1$ and $\tilde{\chi}^{0}_2 \rightarrow h(\to WW^*\,{\rm or}\,ZZ^*)\tilde{\chi}^{0}_1$. 
To obtain generic results for a Higgs final state, the collaboration assumed SM branching ratios for the $h$, 
with the mass fixed to $m_h=125$ GeV. 
To use these results it is important to identify the SM-like Higgs for any given parameter point. 
When calling {\tt  smodels()}, a check is performed on all neutral scalar particles with a mass in the range 123--128~GeV, 
comparing their branching ratios to $WW,ZZ,\tau\tau, b\bar{b}$ to those of a SM Higgs of the same mass. 
If they are compatible within $15\%$, the corresponding particle will be identified as a SM Higgs 
by an entry of type
\begin{verbatim}
25 : "higgs",
-25 : "higgs"
\end{verbatim}
in the {\tt rEven} dictionary in the file {\tt particles.py}.    
Note that the name {\tt higgs} is reserved for a SM-like Higgs and should not be assigned generically. 
If no particle of that name is identified in {\tt particles.py}, \smodels\ assumes that there is no SM-like Higgs, 
and the corresponding SMS results requiring a Higgs in the final state are not used to constrain the parameter point.
However, for internal consistency, the name ``higgs'' has to be defined, and will be assigned to the PDG ID 12345.

When directly submitting an SLHA input file without calling {\tt smodels()}, one has to
make sure that a correct particles.py exists in the directory from which \micro\ is run (the directory where \verb|main.c| is located). For details on the syntax, see~\cite{Kraml:2014sna,Ambrogi:2017neo}.

Note that the branching ratios of the SM Higgs are computed within \micro, while the branching ratios in the new physics model are by default read from the SLHA input file. Differences in computing decays of the Higgs into off-shell particles or in the choice of fundamental constants can lead to the misidentification of the SM-like particle. 
If this occurs it is safer to simply  disable the readout of the decay tables, such that all branching ratios will be consistently calculated by \micro, see~\cite{Belanger:2013oya}.

\subsubsection{SMS caveats}
\label{sec:caveats}

A set of assumptions is introduced when defining the generic description of SMS.
First, the production channel is not taken into account, and only on-shell particles are considered in the cascade decay.
Virtual particles are replaced by an effective vertex, where only the on-shell decay products are specified.
Additionally, new states are described only by their mass, neglecting all other quantum numbers, while in general different spin structures might modify selection efficiencies.
Finally it should be noted that the SMS approach is only valid within the narrow width approximation.
For a safe application of \smodels\ (in particular to non-MSSM scenarios), the above mentioned assumptions should be understood and if needed verified. 
The validity of these assumptions will depend on the concrete model under consideration, as well as details of the experimental search. In particular, an inclusive cut-and-count search might be less sensitive to differences than a shape-based analysis or a multivariate analysis.

Previously, the effects of alternative production channels in squark simplified models were studied in~\cite{Edelhauser:2014ena}, the effect of a different spin structure for the case of the dijet+missing transverse energy (MET) final state was studied in~\cite{Edelhauser:2015ksa}, and the effect of a different spin structure for the dilepton+MET final state was studied in~\cite{Arina:2015uea}.
Recently the spin dependence in $t\bar{t}$+MET final states was tested in~\cite{Kraml:2016eti}. 
For all these cases it was found that the application of SMS limits is safe.

When the underlying assumptions that enter the simplified models interpretation are  too restrictive to probe some parameter space of a model,  more comprehensive  recasting codes such as \checkmate~\cite{Drees:2013wra,Kim:2015wza} or \madanalysis~\cite{Dumont:2014tja,Conte:2014zja}, should be used instead. These codes however  require generating events for the new physics signal and are therefore more computer-time consuming and less adapted to large scans. 
Recasting by event simulation is not directly interfaced in this distribution. 

\subsection{Other simplified-model limits}

\subsubsection{\boldmath $Z'$ searches}

Limits on a new massive Abelian gauge boson from various searches at the LHC are taken into account in \micro\ through a routine originally designed for the UMSSM model~\cite{Belanger:2011rs,Belanger:2015cra} but which can be adapted to other models with a $Z'$ uniquely defined by the PDG code 32. The usage is 
\begin{quote}
  \verb|Zprimelimits()|
\end{quote}
which returns 0 if the point in the parameter space of the model is not excluded by the $Z'$ constraints, 1 if the point is excluded and 2 if both subroutines dealing with $Z'$ constraints cannot test the given scenario.

Currently two types of searches defined in different subroutines of \verb|Zprimelimits()| are implemented.
The latest $Z'$ search in the dilepton final state at $\sqrt{s} = 13$~TeV from ATLAS~\cite{Aaboud:2016cth} is considered in the first subroutine \verb|Zprime_dilepton|.  
It excludes, for instance, a Sequential Standard Model (SSM) $Z'$ up to $3.36$~TeV. 
The evaluation done by ATLAS however assumes that the $Z'$ only decays into SM particles. 
The limit can therefore be relaxed if the $Z'$ also has decay modes into new particles, 
thus reducing the branching ratio into dileptons. 
For this reason, after the cross section $\sigma(pp\rightarrow Z')$ at $\sqrt{s} = 13$~TeV is computed with the \texttt{hCollider} function of \micro, a rescaling factor is applied to  
  $\sigma(pp\rightarrow Z')\times {\rm BR}(Z'\rightarrow l^+l^-)$, where $l^+l^-$ stands for the combined dielectron and dimuon channels,  to match the computed cross section in the limit where $Z'$ only decays into SM particles  used by ATLAS.
This is then compared  with the limit set by ATLAS in the range $M_{Z'} \in [0.5; 4]$~TeV. 
The subroutine returns~1 if $\sigma(pp\rightarrow Z')\times {\rm BR}(Z'\rightarrow l^+l^-)$ exceeds the observed limit. 

If the scenario considered is allowed or not tested by \verb|Zprime_dilepton|, a second subroutine called \verb|Zprime_dijet| analyses the point using constraints from LHC dijet searches at $\sqrt{s} = 8$~TeV~\cite{Aad:2014aqa,Khachatryan:2015sja,Khachatryan:2016ecr} and at $\sqrt{s} = 13$~TeV~\cite{ATLAS:2015nsi,Khachatryan:2015dcf}.
This subroutine uses the recasting  performed  in \cite{Fairbairn:2016iuf} for  a  combination of ATLAS and CMS searches. The recasting provides an upper bound at 95\% CL on $g_q^2 \times \br(Z' \rightarrow jj)$ where $g_q$ is the coupling of $Z'$ to quarks $u,d,c,s$ or $b$ for  set of values for \{$M_{Z'}$, $\Gamma/M_{Z'}$\} where $\Gamma$ is the total $Z'$ width.  Only quarks (except top) and invisible particles are included in the calculation of the $Z'$ width. For the relevant \{$M_{Z'}$, $\Gamma/M_{Z'}$\}, \verb|Zprime_dijet| compares this upper bound   with the value of $\sum_q \left[ \left(g^V_q\right)^2 + \left(g^A_q\right)^2 \right] \times \br(Z' \rightarrow q\bar{q})$ corresponding to the scenario considered : it returns 1 if the result exceeds the upper bound and 0 otherwise. Here $q=u,d,c,s,b$; $g^V_q$ and $g^V_q$ are the vectorial and axial couplings of $Z'$ to $q$ which have to be defined in the model files. If $M_{Z'}>4$~TeV or no coupling is defined this subroutine cannot be used  and returns automatically 2.
Note finally that \verb|Zprimelimits()| returns 1 if $M_{Z'}<0.5$~TeV and  2 for points  for which the narrow-width approximation is not valid, \ie~$\Gamma/M_{Z'}>$~0.3.

\subsubsection{Mono-jet searches}

At the LHC, DM searches mainly proceed via mono-$X$ signatures, where $X$ indicates any visible, \ie\ collider detectable, particle produced in association with a DM pair. These signatures are characterised by a high-$p_T$ object recoiling against MET. Currently, the most stringent limits arise from the 8~TeV mono-jet searches~\cite{Khachatryan:2014rra,Aad:2015zva}, while early 13~TeV results are quickly reaching the Run-1 sensitivity~\cite{CMS:2015jdt}.  The function\\

   \verb|monoJet(pName1,pName2)|\\
   
\noindent
computes the cross section for $p,p\rightarrow$ \verb|pName1,pName2|$ +jet$ at $\sqrt{s}=8$~TeV 
where \verb|pName1|, \verb|pName2| are the names of neutral outgoing particles and $jet$ includes light quarks (u,d,s)  and gluons.
The cross section is computed through a built-in map that takes into account the differences between parton level and detector reconstructed events as well as next to leading order corrections to the signal prediction.\footnote{We have used the NLO DM model publicly available at the FeynRules~\cite{Alloul:2013bka} page \url{http://feynrules.irmp.ucl.ac.be/wiki/DMsimp}.} This map has been created via the {\tt MadGraph5\_aMC@NLO}~\cite{Alwall:2014hca}, {\tt PYTHIA8}~\cite{Sjostrand:2007gs} and {\tt CheckMate}~\cite{Drees:2013wra} chain for various MET selections and mediator masses. The function convolutes the parton level prediction with this map to compute the final rates for each signal region (SR) of the CMS analysis~\cite{Khachatryan:2014rra}, which are then compared with the experimental background and data.
The function returns  the resulting  CL  obtained with the CLs technique~\cite{Read:2002hq,Read:2000ru} for each  SR and chooses the most constraining one. We checked that in the case of a vector mediator, the exclusion levels match those provided by CMS.

\section{Installation and sample output}
\label{sec:install}

The code can be downloaded from~\cite{micro-hp}.
After unpacking, go to the \verb|micromegas_4.3.4| directory and type\\

   \verb|make| ~or~ \verb|gmake|\\

\noindent
To work with one of the models already included in the distribution, go to the relevant model directory (for example MSSM) \\

\verb|cd MSSM|\\

\noindent
The file {\tt main.c} is a sample program to illustrate the usage of the functionalities described in this paper. 
It contains at the top of the code various options to steer the behaviour of the code. For example  

\begin{verbatim}
#define RGE suspect
#define EWSB
#define MASSES_INFO
#define CONSTRAINTS
#define HIGGSBOUNDS
#define HIGGSSIGNALS
#define LILITH 
#define SMODELS 
\end{verbatim}
will compute the MSSM spectrum using \suspect\ with input parameters specified at the electroweak scale (to be provided in an input file, for example {\tt mssms.par}, as described in \cite{Belanger:2001fz}). The resulting mass spectrum will be displayed on the screen together with various LEP  and flavour constraints, including the ones discussed in Section~\ref{sec:lep}. In this example, the Higgs constraints will be checked by \HB\ and \HS\  as well as with \lilith.  Note, however, that the latter two perform the same task and only one should be used in, e.g., a global fit.
The LHC constraints on the SUSY spectrum will be checked with \smodels.\footnote{Running \smodels\ requires computing production cross sections for all non standard particles in the model. The compilation of these processes at the first call of \micro\ may take a few minutes.}

The codes \suspect,  \lilith\ and  \smodels\ are distributed with \micro\ and are located in the \verb|micromegas_4.3.4/Packages| directory. The codes \HB\ and \HS, when requested, will be copied automatically from a repository upon compilation of \micro. 
They will then also be stored in the \verb|micromegas_4.3.4/Packages| directory. 

Compiling and running {\tt main.c} with the above options and the input file \verb|mssms.par| 
\begin{verbatim}
   make main=main.c
   ./main mssms.par
\end{verbatim}
will produce the following output
\footnotesize{
\begin{verbatim}
========= EWSB scale input =========
 Spectrum calculator is suspect
Initial file  "mssms.par"
The following parameters keep default values:
  alfSMZ=1.1840E-01      MW=8.0200E+01      MZ=9.1187E+01      Ml=1.7770E+00
    McMc=1.2700E+00    MbMb=4.2300E+00      Au=0.0000E+00      Ad=0.0000E+00
 SU_read_leshouches: end of file
 OUTPUT in SLHA format in suspect2_lha.out 
 RUN TERMINATED : OUTPUT in suspect2.out
Warnings from spectrum calculator:
 .....none

Dark matter candidate is '~o1' with spin=1/2  mass=1.89E+02

~o1 = 0.950*bino -0.062*wino +0.256*higgsino1 -0.167*higgsino2

=== MASSES OF HIGGS AND SUSY PARTICLES: ===
Higgs masses and widths
      h   125.38 4.52E-03
      H   699.89 5.23E+00
     H3   700.00 5.85E+00
     H+   704.60 4.81E+00

Masses of odd sector Particles:
~o1  : MNE1  =   188.5 || ~l1  : MSl1  =   198.6 || ~eR  : MSeR  =   204.8 
~mR  : MSmR  =   204.8 || ~1+  : MC1   =   283.4 || ~o2  : MNE2  =   289.5 
~o3  : MNE3  =   310.3 || ~2+  : MC2   =   441.8 || ~o4  : MNE4  =   442.0 
~ne  : MSne  =   496.0 || ~nm  : MSnm  =   496.0 || ~nl  : MSnl  =   496.0 
~eL  : MSeL  =   502.0 || ~mL  : MSmL  =   502.0 || ~l2  : MSl2  =   502.6 
~t1  : MSt1  =  1369.9 || ~b1  : MSb1  =  1552.5 || ~uL  : MSuL  =  1557.9 
~cL  : MScL  =  1557.9 || ~uR  : MSuR  =  1558.3 || ~cR  : MScR  =  1558.3 
~dR  : MSdR  =  1559.0 || ~sR  : MSsR  =  1559.0 || ~dL  : MSdL  =  1559.8 
~sL  : MSsL  =  1559.8 || ~b2  : MSb2  =  1566.3 || ~t2  : MSt2  =  1656.3 
~g   : MSG   =  1842.8 || 


==== Physical Constraints: =====
deltartho=3.44E-04
gmuon=1.12E-09
bsgnlo=3.21E-04 ( SM 3.28E-04 )
bsmumu=3.28E-09
btaunu=9.59E-01
dtaunu=5.17E-02  dmunu=5.33E-03
Rl23=9.998E-01
MassLimits OK
HB(4.3.1): result=1  obsratio=5.15E-01  channel= (pp)->h2->tautau,
           using -2ln(L) reconstruction  (CMS-HIG-PAS 14-029) 
HS(1.4.0): Nobservables=89 chi^2 = 7.74E+01 pval= 8.06E-01
LILITH(DB15.09):  -2*log(L): 28.14; -2*log(L_reference): 0.00; ndf: 38; p-value: 8.79E-01 

SMODELS:
found SM-like Higgs = h
v1.1.0patch1 with database 1.1.0patch1 
highest R=3.16E-01 from ATLAS-CONF-2013-047, topology T2
not excluded.
\end{verbatim}}

\normalsize
The user can also implement his/her own model in \micro, this requires creating model files in the \chep\ format,  see~\cite{Belanger:2006is} for details.
The command 
\begin{verbatim}
./newProject ModelName
\end{verbatim}
will create a directory {\tt ModelName} containing sample main files that include the functions necessary to compute 
all dark matter observables and collider constraints.

\section{Conclusions}
\label{sec:Conclusion}

Version 4.3.1 of \micro\ allows for a fast an efficient exploitation of the collider results on new particle searches and on Higgs properties. This is achieved through interfaces with \lilith~\cite{Bernon:2015hsa} and \HS~\cite{Bechtle:2013xfa} 
for checking the Higgs signal strengths, \HB~\cite{Bechtle:2011sb,Bechtle:2013wla} to check limits on additional neutral and charged Higgs states, and \smodels~\cite{Kraml:2013mwa,Kraml:2014sna,Ambrogi:2017neo} to check the LHC limits on the production of new odd particles in the context of Simplified Model Spectra constraints. In addition, in-house routines are provided for checking LEP constraints, limits on a new $Z'$ at the LHC, and LHC limits from mono-jet searches. 
For each of these codes, we explained the basic functionality, the usage within \micro\, the input format and the interpretation of the output. 

\micro\ thus provides a general framework for evaluating collider constraints on dark matter models. 
The framework is fast enough to be suitable for scans of parameter space. It is also quite generic and thus applicable to a wide range of models, including new ones that can be added to \micro. Of course some limitations apply. For example, the Higgs 
signal strength approach in \lilith\ and \HS\ applies only to Higgs sectors with the same tensor structure and Higgs production modes as in the SM. Moreover, the SMS approach used in \smodels\  is subject to a few caveats explained in the paper. 
Finally, the $Z'$ limits are applicable only to models with one extra $Z$ boson. The user is responsible for 
observing  these limitations.

\section{Acknowledgements}

We thank Tim Stefaniak for discussions on \HS. 
This work was supported in part by the LIA-TCAP of CNRS, by the French ANR, Project DMAstro-LHC ANR-12-BS05-0006, 
the Theory-LHC-France Initiative of INP/IN2P3, the {\it Investissements d'avenir}, Labex ENIGMASS, and by the Research Executive Agency (REA) of the European Union under the Grant Agreement PITN-GA2012-316704 (``HiggsTools").
The work of AP  was  also supported by the Russian foundation for Basic Research, grant RFBR-15-52-16021-CNRS-a.


\providecommand{\href}[2]{#2}\begingroup\raggedright\endgroup

\end{document}